\documentclass[preprint,times,3p,twocolumn]{elsarticle}
\usepackage{hyperref}
\usepackage{amsmath}
\usepackage{subfigure}

\usepackage{docmute}

\makeatletter
\def\ps@pprintTitle{%
 \let\@oddhead\@empty
 \let\@evenhead\@empty
 \let\@oddfoot\@empty
 \let\@evenfoot\@oddfoot
}
\makeatother

\begin{document}
\begin{frontmatter}
\title{2-mm-Thick Large-Area CdTe Double-sided Strip Detectors for High-Resolution Spectroscopic Imaging of X-ray and Gamma-ray with Depth-Of-Interaction Sensing}
\author[ut,kavli]{Takahiro Minami\corref{cor1}}
\author[kavli]{Miho Katsuragawa}
\author[ut,kavli]{Shunsaku Nagasawa}
\author[kavli,imagineX]{Shin'ichiro Takeda}
\author[isas,kavli]{Shin Watanabe}
\author[riken]{Yutaka Tsuzuki}
\author[kavli,ut]{Tadayuki Takahashi}
\address[ut]{Department of Physics, University of Tokyo, 7-3-1 Hongo, Bunkyo, Tokyo 113-0033, Japan}
\address[kavli]{Kavli Institute for the Physics and Mathematics of the Universe (Kavli IPMU, WPI), University of Tokyo5-1-5 Kashiwanoha, Kashiwashi, Chiba 277-8583, Japan}
\address[isas]{Institute of Space and Astronautical Science, Japan Aerospace Exploration Agency (ISAS/JAXA), 3-1-1 Yoshinodai, Chuo-ku, Sagamihara, Kanagawa 252-5210, Japan}
\address[imagineX]{iMAGINE-X Inc., 604 ILA Shibuya Mitake Building, 1-12-8 Shibuya, Shibuya-ku, Tokyo 150-0002, Japan}
\address[riken]{Nishina Center, RIKEN, 2-1 Hirosawa, Wako, Saitama 351-0198, Japan}
\cortext[cor1]{Email Address: takahiro.minami@ipmu.jp}

\begin{abstract}
We developed a 2-mm-thick CdTe double-sided strip detector (CdTe-DSD) with a 250 $\mu$m strip pitch, which has high spatial resolution with a uniform large imaging area of 10 cm$^2$ and high energy resolution with high detection efficiency in tens to hundreds keV. The detector can be employed in a wide variety of fields for quantitative observations of hard X-ray and soft gamma-ray with spectroscopic imaging, for example, space observation, nuclear medicine, and non-destructive elemental analysis.
This detector is thicker than the 0.75-mm-thick one previously developed by a factor of $\sim$2.7, thus providing better detection efficiency for hard X-rays and soft gamma rays.
The increased thickness could potentially enhance bias-induced polarization if we do not apply sufficient bias and if we do not operate at a low temperature, but the polarization is not evident in our detector when a high voltage of 500V is applied to the CdTe diode and the temperature is maintained at -20 $^\circ$C during one-day experiments.
The “Depth Of Interaction” (DOI) dependence due to the CdTe diode's poor carrier-transport property is also more significant, resulting in much DOI information while complicated detector responses such as charge sharings or low-energy tails that exacerbate the loss in the energy resolution.

In this paper, we developed 2-mm-thick CdTe-DSDs, studied their response, and evaluated their energy resolution, spatial resolution, and uniformity. We also constructed a theoretical model to understand the detector response theoretically, resulting in reconstructing the DOI with an accuracy of 100 $\mu m$ while estimating the carrier-transport property. 
First, we evaluated the energy resolution using $^{22}$Na, $^{133}$Ba, and $^{57}$Co reconstructing the DOI effect from the energy correlation on both the anode and cathode sides. We obtained an energy resolution of 4.3 keV (FWHM) at 356 keV.
Second, we formulated a theoretical detector-response model that reproduced the experimental data. Using the model, we determined the mobility-lifetime products of carriers $\mu\tau_{e,h}=(4 \pm 1 )\times 10^{-3}, (1.15\pm 0.05) \times 10^{-4}$ [cm$^2$/V] with the space charge density of $n_{sp}=-6.0\times 10^{10}$ [1/cm$^{-3}]$ and then reconstructed the DOI with an accuracy of 100 $\mu m$. 
Finally, we evaluated the spatial resolution and demonstrated to resolve patterns of 250-$\mu$m-width slits with an X-ray test chart. We found no positional variation in the detector response. 
We realized the detector that has high energy resolution and high 3D spatial resolution with a uniform large imaging area.

\end{abstract}

\end{frontmatter}

\section{Introduction}
Spectroscopic imaging in tens to hundreds of keV are ever more extensively used in recent decades in a variety of research fields,
such as X-ray astronomy \cite{astro-h,Harrison_2013}, non-destructive elemental analysis \cite{osawa2023}, and in vivo imaging in nuclear medicine \cite{takeda2018high}.
Cadmium telluride (CdTe) is one of the attractive elements for hard X-ray and soft gamma-ray detectors owing to the generally high atomic numbers and density of the detectors of these types \cite{takahashi2001recent}.
In addition, high-purity CdTe single crystals with a diameter of 5 cm were produced by Acrorad Co., Ltd., using the Traveling Heater Method \cite{FUNAKI1999120}. The recent development of CdTe Schottky diodes achieved very low leakage current and high energy resolution \cite{takahashi1999high,Toyama_2004,ABBENE2013135}.

CdTe Double-sided Strip Detectors (CdTe-DSDs) with the CdTe Schottky diodes \cite{watanabe2009} were originally developed for astrophysical applications \cite{TAKAHASHI2005next,lindsayFOXSI} and are employed in various fields \cite{katsuragawa2018compact,Yagishita2022,Fukuchi_2023}. 
Notably, the CdTe-DSD with a large area and high energy and spatial resolution \cite{kokubun_hxi} is suitable in medical imaging, in particular, to observe small animals in combination with parallel-hole collimators, enabling multi-isotope imaging and identification of small organs and tissues \cite{Fujiipinhole}.
However, the previously developed CdTe-DSD with a thickness of 750 $\mu$m has insufficient detection efficiency for energies above 80 keV.

An intuitive method to improve the detection efficiency is to employ thicker CdTe diodes; however, there is a potential downside in the energy and spatial resolution. 
First, the CdTe is known to have a relatively poor transport property of carriers, especially holes, which causes the dependence of the charge collection efficiency of the detector on the ``Depth Of Interaction'' (DOI). The dependence increases in the thicker CdTe diodes, resulting in more significant low-energy tails that exacerbate the loss in energy resolutions \cite{takahashi2001recent}.
Second, the direction of incident photons cannot be evaluated with sufficient accuracy for specific applications if the position of DOI in the thick detector is not determined. The angular resolutions of imaging systems, such as SPECT, Compton camera, and coded aperture camera, could be exacerbated if the direction is uncertain \cite{SPECT_Meikle}.
Recently, the analysis method of reconstructing the energy spectrum and the DOI position for CdTe or CdZnTe detectors \cite{furukawa2020imaging, salccin2014fisher,BOLOTNIKOV201641,VANPAMELEN1998390,KimJaeCheon2014,7908933,6304947,VERGER200733} for the previous CdTe-DSDs was reported and can be clues to solving the downside. 

Here we present the 2-mm-thick CdTe-DSD that we have developed with the aim of improving the detection efficiency over 80 keV (Section 2) and the method that we implemented for reconstructing the energy spectrum, and we evaluated its energy resolution (Section 3). Further, we also developed a simulated model of the detector to understand its response and verified its reliability by comparing the model results with experimental data. By using this model, we build a new method to reconstruct the DOI (Section 4). The spatial resolution and the potential dependence of the detector response on the detected position were investigated (Section 5).
These new methods could be applicable to thick double-sided strip detectors of CdTe or CdZnTe in general.

\begin{figure}[ht!]
\begin{center}
\includegraphics[width=0.8\hsize]{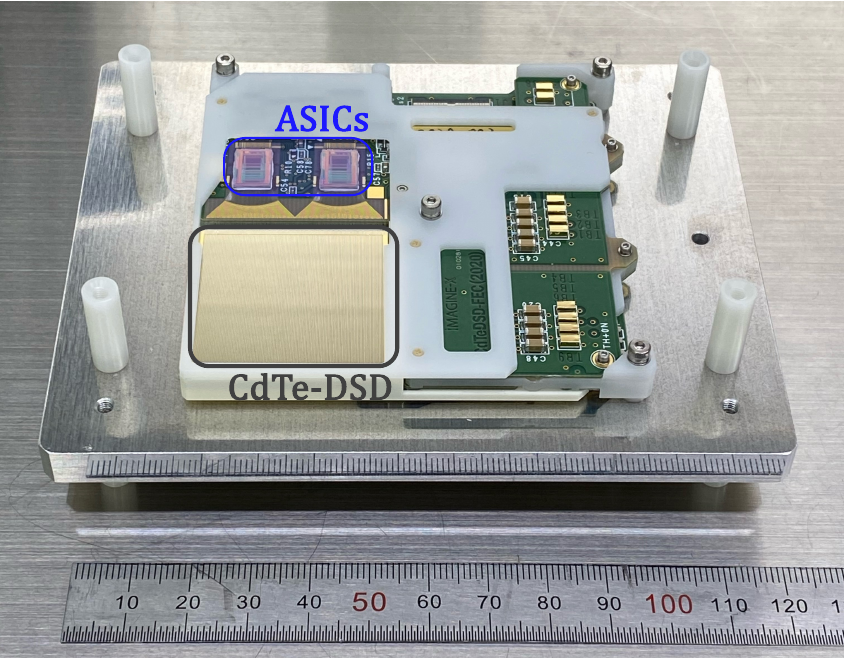}
\includegraphics[width=0.8\hsize]{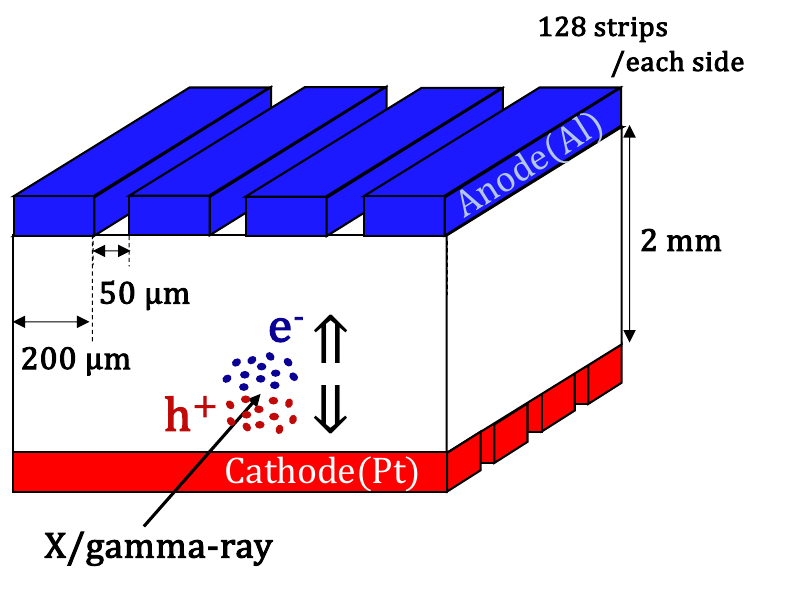}
\caption{(upper) Photograph of the 2-mm-thick CdTe-DSD and (lower) schematic view of the electrode configuration. The detector is connected to a VATA-SGD ASIC 
\cite{WATANABE2014192} for pulse-height measurement. Electrons and holes generated by incident photons travel along the electric field to the anode and cathode electrodes, respectively. }
\label{fig:Detector_config_fig}
\end{center}
\end{figure}

\section{Detector configurations}
Figure \ref{fig:Detector_config_fig} shows a photograph and schematic view of the 2-mm-thick CdTe-DSD we developed. The detector has 128 strip electrodes on each side. 
The anodic strips are orthogonal to the cathodic ones.
Each strip has a width of 200 $\mu$m and is placed with a gap of 50 $\mu$m from the immediately adjacent strips. 
The imaging area is 32 $\times$ 32 mm$^2$ with 16384 pixels. The cathode and anode electrodes are made of platinum (Pt) and aluminum (Al), respectively, and are powered with voltages of 0 and 500 V, respectively.
The DSD configuration enabled us to acquire energy information on both the electrodes (anode and cathode sides). 

\begin{figure*}[ht!]
\centering
\includegraphics[width=0.8\hsize]{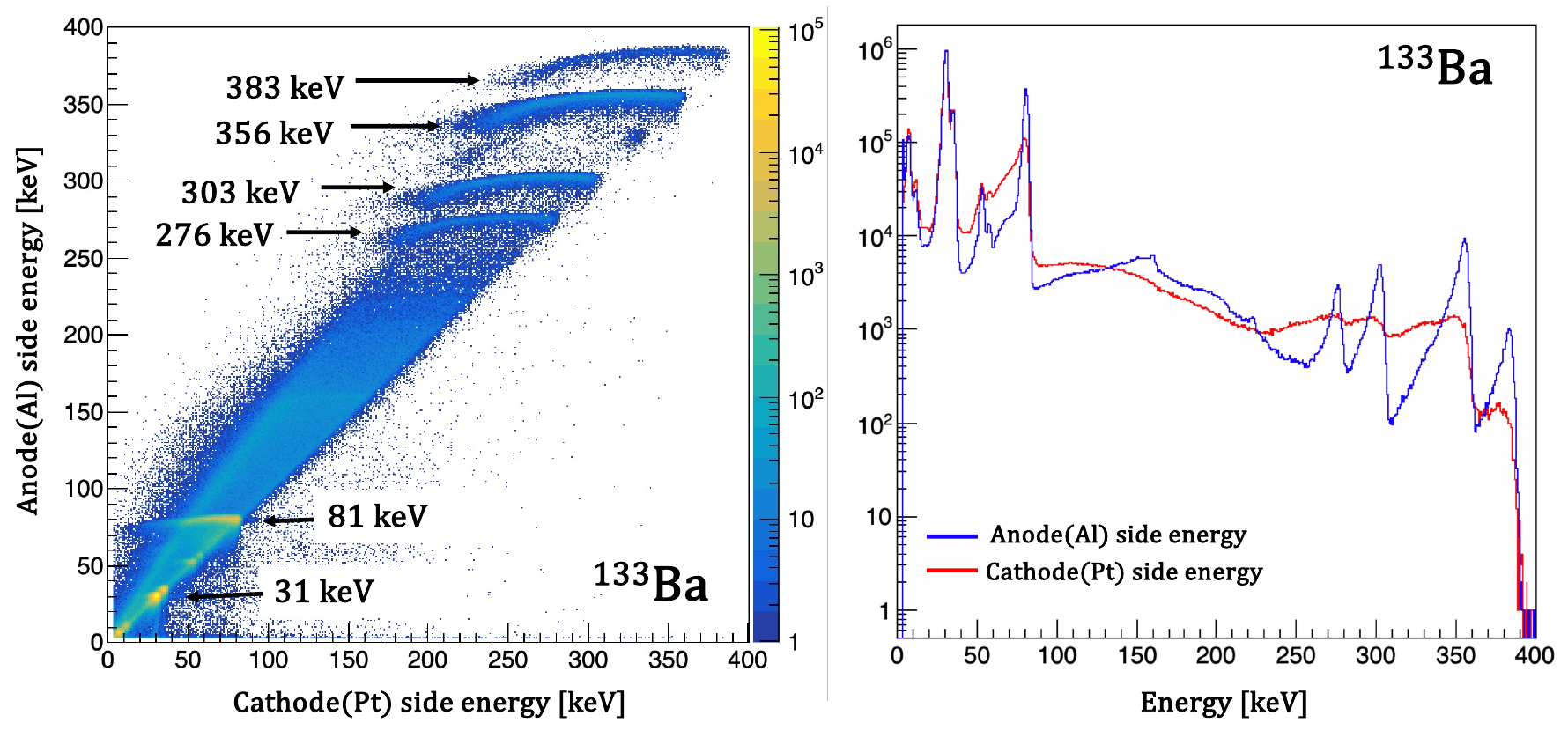}
\caption{(left) Correlation of the detected energies of $^{133}$Ba between the anode and cathode sides for single-strip events. (right) Spectra of the cathode (red line) and anode-side (blue line) events.
 Each energy peak shows a significant tail toward the lower energy on the cathode side.}
\label{fig:Corerala+both_side_spectrum_133Ba-2}
\end{figure*}

 When the CdTe interacts with X/gamma rays, pairs of electrons and holes are produced and move to the anode and cathode, respectively. Thus, the detected energies on both sides and the two-dimensional X/gamma-ray interaction position are calculated using the information on the pulse height and position of the strips on which the signal is detected. 

The DSD configuration also helped us realize compact readout in combination with specially designed Application Specific Integrated Circuits (ASICs) \cite{WATANABE2014192}
The CdTe and ASIC for pulse-height measurement are placed on a front-end card (FEC).
The strip electrode and the ASIC are DC-coupled with wire bonding for the cathode side and bump bonding for the anode side. To ensure accurate signal transmission from the detector, we use a method of floating bias operation, which was originally established for the Hard X-ray Imager of the Hitomi (ASTRO-H) satellite \cite{SATO2016235}. The ASICs on each side are mounted separately in areas isolated by digital isolators on the FEC. When we supply a voltage between the isolated areas on the FEC, a bias voltage is applied to the CdTe-DSD.

The CdTe-DSD is equipped with four ASICs (VATA-SGD \cite{WATANABE2014192}). Each channel of each ASIC has a Charge Sensitive Amplifier (CSA) and two CR shaping circuits, which shape the pulse amplified with the CSA with a fast time constant ($\sim$ 0.6 $\mu$s) to form a trigger signal and with a slow time constant ($\sim$ 3 $\mu$s) to acquire the height of the pulse. After an appropriate fixed time delay from the trigger signal that we determined experimentally to hold the peak of the pulse with a slow time constant, the holding signal is generated, and the pulse height shaped with a slow time constant is sampled, held, and output.

\section{Spectroscopic Performance}
\subsection{Cathode (Pt) and Anode (Al) Spectra and Their Correlation}
To evaluate the detector's spectroscopic performance, we measured the energy spectra using radioactive sources of $^{22}$Na, $^{133}$Ba, and $^{57}$Co with an operating temperature of about $-$20 $^\circ$C. 
In the analysis, the event threshold for each strip on the anode and the cathode side was set to 3$\sigma_\mathrm{{ped}}$ ($\sim$ 3 keV) of each pedestal width. 
Events in which the signal value of only one strip out of all strips on one side exceeded their event threshold value were selected and used for energy calibration of each channel. We call those events ''single-strip events''. By using those events, each energy calibration function for each channel was created to convert the digital value of the pulse height to keV using each peak of the radioisotopes in the spectrum and applied to all events.
We called the calibrated value in units of keV ``anode or cathode-side energy'' for single-strip events. When the signals over the analytical threshold were detected in two or more (adjacent) strips on one side, we regarded the sum of the energies on all the adjacent strips as the ``anode or cathode-side energy'' of the event and called it an ``multi-strip event''.

The detected energies obtained through independent energy measurements on both sides depend on the DOI. Figure \ref{fig:Corerala+both_side_spectrum_133Ba-2} shows the correlation plot between the cathode-side energy $E_\mathrm{cathode}$ and the anode-side energy $E_\mathrm{anode}$ for $^{133}$Ba events detected on only one strip on the anode side and its spectra of the anode and cathode sides. 
The correlation plot (left panel) clearly shows a relationship of detected energies on both sides for a specific incident photon energy, the structure of which extends roughly horizontally (i.e., along the $E_\mathrm{cathode}$ axis) and is slightly curved in the vertical direction (i.e., along the $E_\mathrm{anode}$ axis). 
The broadness of these structures corresponds to the broadness of each peak in the spectra (Figure~\ref{fig:Corerala+both_side_spectrum_133Ba-2} right panel), characterized with a low-energy tail. The structure is also a visualization of the fact that the spectral peaks at the cathode side are much broader than those at the anode side. A closer look finds that the higher the incident photon's energy is, the more extended the structure is. This fact means that the low-energy tail is more pronounced at the higher incident photon energy, resulting in a poorer energy resolution.

\subsection{Incident energy dependence on the number of simultaneously detected strips} 
\begin{figure}[ht!]
\begin{center}
\includegraphics[keepaspectratio,width=0.95\hsize]{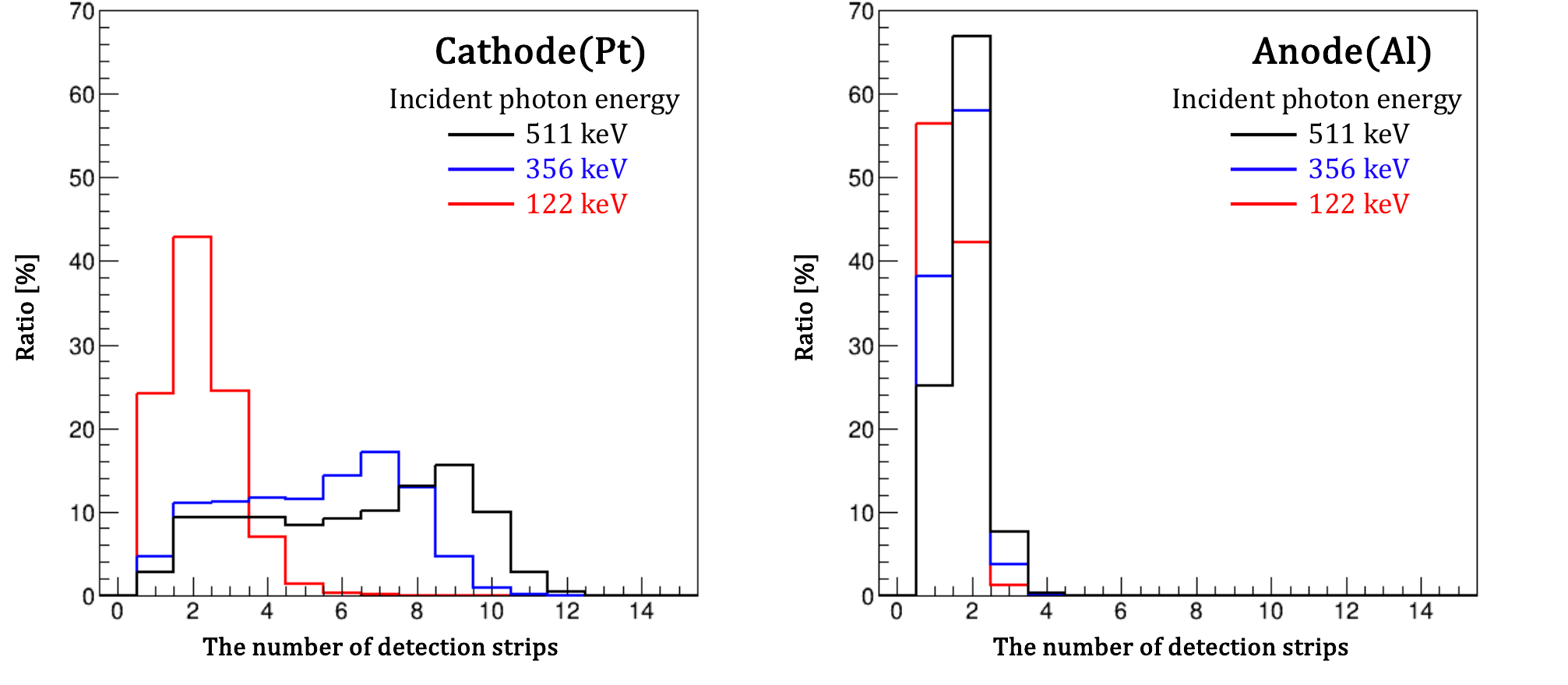}
\caption{Distribution of the number of strips that detected signals at the (left) cathode (Pt) side and (right)  anode (Al) side for incident photons, whose energies are 511 keV ($^{22}$Na, black line), 356 keV ($^{133}$Ba, blue line) and 122 keV($^{57}$Co). }
\label{fig:the_number_of_detection_strip_122+356+511}
\end{center}
\end{figure}

The distributions of multi-strip events vary with the energy of the incident photon and also differ between the anode and cathode sides.
Figure \ref{fig:the_number_of_detection_strip_122+356+511} shows the distributions of single or multiple strip events (n-strip events) for 511 keV\ ($^{22}$Na), 356 keV\ ($^{133}$Ba), and 122 keV\ ($^{57}$Co).
The distribution on the anode side is dominated by 1-strip and 2-strip events, which account for more than 90\% of the distribution at 511 keV and below, with the fraction of 2-strip events increasing as energy increases. This trend can be attributed to the larger size of the charge cloud for higher energies \cite{furukawa2020imaging}. By contrast, the distribution on the cathode side is much broader, especially at 356 and 511 keV.

To investigate the cause of the broad distribution in the cathode, we examined the distributions of the energy deposits at the cathode electrode for individual multi-strip events. The distribution had a high-deposit central peak spreading for only a couple of strips and a much lower-deposit, broad part spreading over peripheral strips. The central strips in the energy distribution of the cathode accounted for $\sim$ 90\% of the overall deposited energy, while the peripheral strips account for only about 10\%.
Hence, considering that the distribution on the anode side has a peak at two strips, the energy deposits on the cathode side and the distribution of the anode side are consistent. 
We conjecture that the central peak in energy deposits at the cathode electrode is determined by the initial charge cloud size and the thermal diffusion of carriers, and the spreading of the charges over the peripheral part mainly from the hole trapping phenomena rather than the thermal diffusion of carriers.

\subsection{Reconstruction of incident photon energies} 
To obtain good resolution over a wide energy range, we need to accurately estimate the relationship between each point in the correlation map shown in Figure \ref{fig:Corerala+both_side_spectrum_133Ba-2} and incident photon energy.
Having noticed that the extended structure generated by a given energy incident on the single anode strip events follows a certain curve (see Figure \ref{fig:Corerala+both_side_spectrum_133Ba-2}), we devised an empirical function to model the curve because creating theoretical functions to replicate the structures across all energy bands is challenging. The function was given by
\begin{equation}
\label{eq:counterline_function}
\begin{split}
E_\mathrm{ave} &= a_\mathrm{E_{0}}\\
&+a_\mathrm{quad}( E_\mathrm{diff/2} -a_\mathrm{center})^2\\
&+a_\mathrm{hyp-s}\sqrt{(E_\mathrm{diff/2}-a_\mathrm{center})^2+a_\mathrm{hyp-c}},
\end{split}
\end{equation}
which the $E_\mathrm{diff/2}$ and $E_\mathrm{ave}$ are half of the difference of the detected energies at the cathode and anode sides and the average of those energies, respectively. 
Each term of the function is, in order, a constant, a quadratic function representing a curve of the structure, and a hyperbolic function representing a straight line of the structure.
The $a_\mathrm{E_{0}}$, $a_\mathrm{quad}$, $a_\mathrm{hyp-s,c}$, and $a_\mathrm{center} (\sim 0)$ are fitting parameters. We also found that the structures in events detected on multiple strips on the anode side can be expressed by the above function with different parameter sets.
\begin{figure*}[ht!]
\begin{center}
\includegraphics[width=0.95\hsize]{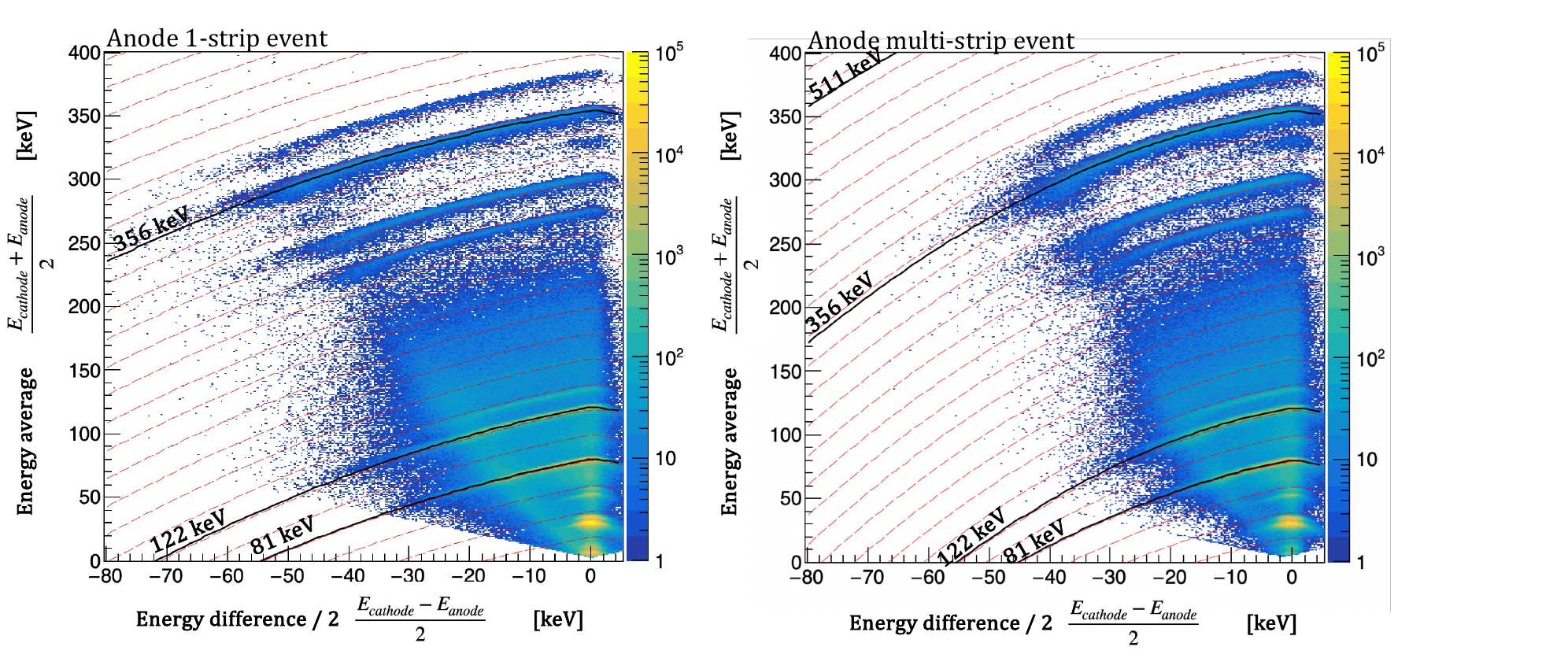}

 \begin{minipage}{0.99\textwidth}
 \centering
 \small
 \begin{tabular}{c|ccccc}
 \multicolumn{6}{c}{Anode single-strip event}\\
   $E_\mathrm{inc}$ [keV] & $a_\mathrm{E_0}$ &
   $a_\mathrm{center}$ &
   $a_\mathrm{quad}$ &
   $a_\mathrm{hyp-s}$ &
   $a_\mathrm{hyp-c}$ \\
   \hline
   81 & 81.09 & 0.50 & -0.01 & -0.91 & 0.65 \\
   122 & 121.84 & 0.50 & -0.01& -0.90 &1.0 \\
   356 & 354.48 & 0.45& -0.01 & -0.72 &0.00 \\
   511 & 509.43 & -0.15& -0.01 &-0.76 &0.00 \\

\multicolumn{6}{c}{}\\
\multicolumn{6}{c}{Anode multi-strip event}\\
   $E_\mathrm{inc}$ [keV] & $a_\mathrm{E_0}$ &
   $a_\mathrm{center}$ &
   $a_\mathrm{quad}$ &
   $a_\mathrm{hyp-s}$ &
   $a_\mathrm{hyp-c}$ \\
   \hline
   81 & 81.18 & 0.26 & -0.02& -0.97 &1.0 \\
   122 & 122.11 & 0.49 & -0.02& -0.86 &1.0 \\
   356 & 354.74 & 0.38& -0.02 & -0.65 &0.00 \\
   511 & 509.77 & -0.10& -0.02 &-0.67 &0.06 \\
  \end{tabular}
 \end{minipage}

\caption{(upper) Correlation of both-side energies with summing the experimental data of $^{133}$Ba and $^{122}$Co in the (upper left) anode single-strip events and (upper right) anode multi-strip events and the contour lines (red dotted line).
The X-axis and Y-axis are an average of the detected energies at the cathode and anode sides $E_\mathrm{ave}=(E_\mathrm{cathode}+E_\mathrm{anode})/2$ and a half of the difference $E_\mathrm{diff/2}=(E_\mathrm{cathode}-E_\mathrm{anode})/2$, respectively. 
The contour lines are calculated from fitted functions $E_\mathrm{ave} = a_\mathrm{E_{0}}+a_\mathrm{quad}( E_\mathrm{diff/2} -a_\mathrm{center})^2+a_\mathrm{hyp-s}\sqrt{(E_\mathrm{diff/2}-a_\mathrm{center})^2+a_\mathrm{hyp-c}}$ with the structures whose incident photon energy $E_\mathrm{inc}$ are 81, 122, 356 and 511 keV (black line) 
The interval of the contour line is 20 keV. 
(bottom) The fitted parameters of the fitted functions with each structure of $E_\mathrm{inc} = $ 81, 122, 356, and 511 keV.}

\label{fig:energy_contourmap}

\end{center}
\end{figure*}

For each of the anode single-strip and anode multi-strip events, we used the function to fit the structures corresponding to an incident photon energy of 81, 122, 356, and 511 keV, and calculated the contour lines using these four fitted functions. Figure \ref{fig:energy_contourmap} shows the correlations plot with summing the experimental data of $^{133}$Ba and $^{122}$Co data, the contour lines, and the fitted values of $a_\mathrm{E_{0}}$, $a_\mathrm{quad}$, $a_\mathrm{hyp-s,c}$, and $a_\mathrm{center}$ for each of the anode single-strip and anode multi-strip events.
The contour lines every 20 keV were created by dividing internally the neighboring fitted functions expressed by equation (\ref{eq:counterline_function}) by the ratio of each incident photon energy.
In addition, the contour lines whose incident photon energy was below 81 keV or above 511 keV were calculated by translating parallelly the fitted function of 81 keV or 511 keV, respectively.

We interpolated between the counter lines with the Delaunay triangulation \cite{Delaunay} to reconstruct the incident photon energies for all data points in the correlation plot. 
 Owing to the limited number of available X- or gamma-ray lines, interpolation is necessary to deduce the response for energies that have not been directly measured. While an analytical 2D response function could be constructed, using interpolation is simpler and allows for a more flexible description of the response changes. 
The Delaunay diagrams are a well-known technique to represent a three-dimensional surface as a triangular mesh. Given that $E_\mathrm{ave}$ and $E_\mathrm{diff/2}$ obtained from experimental data are inside a certain triangle, the interpolated value of the reconstructed energy $E_\mathrm{reconst}$ is easily calculated from an average weighted by the inverse of length in the $E_\mathrm{diff/2}-E_\mathrm{ave}$ plane from each vertex of the triangle. 
Then, the incident photon energies are able to be reconstructed if we know the average of both-side energies and the half of the difference from experimental data.

\begin{figure*}[t]
\begin{center}
\includegraphics[width=0.8\hsize]{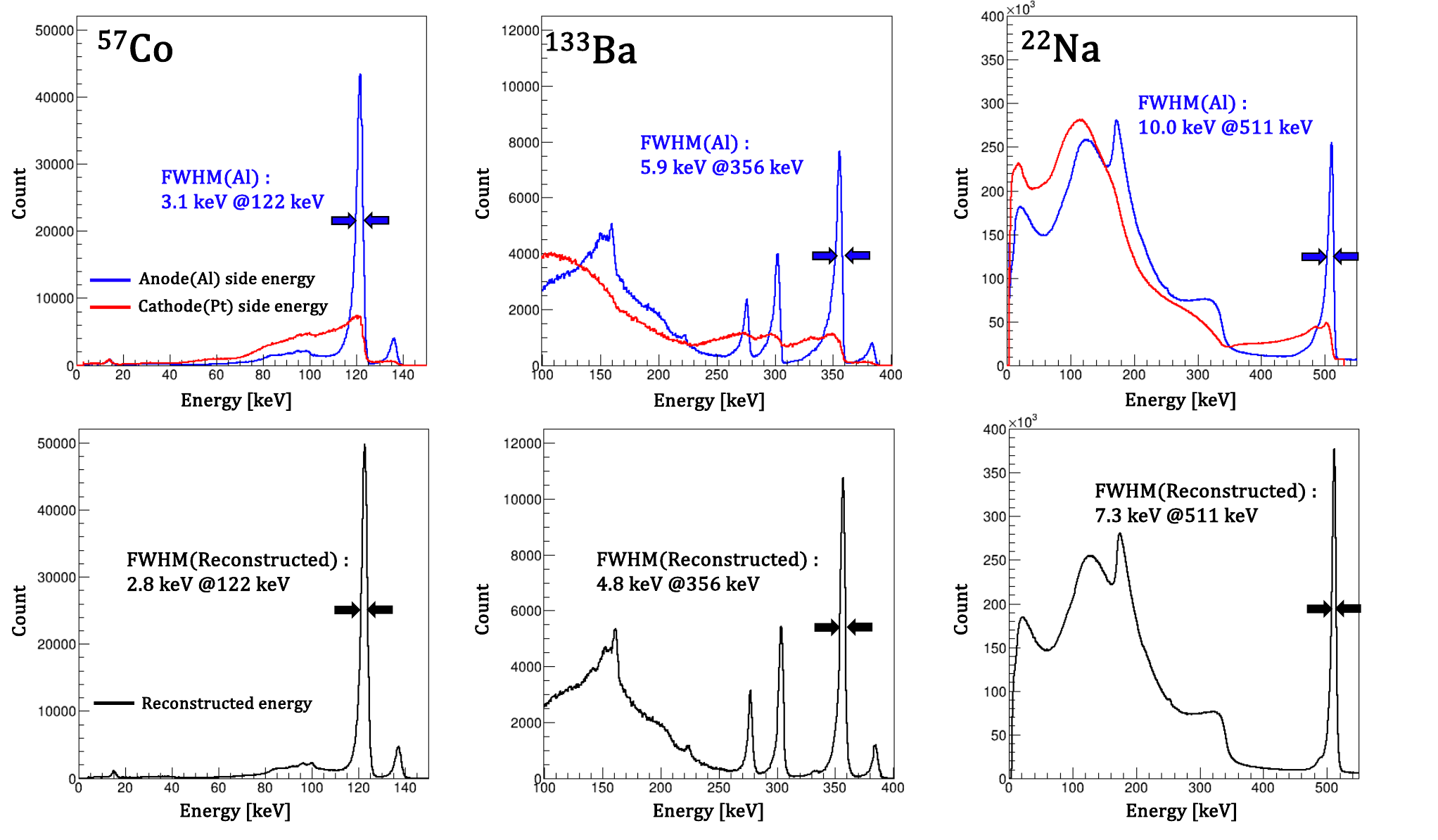}
\caption{(upper\ row) Energy spectra on the cathode side\ (red line) and anode side\ (blue line) and (lower\ row) reconstructed spectra \ (black line) of (left) $^{57}$Co, (center) $^{133}$Ba, and (right) $^{22}$Na, respectably. We used the single and multi-strip events. }
\label{fig:reconst-spectrum-Ba+Na+Co}
\end{center}
\end{figure*}

Figure \ref{fig:energy_contourmap} also shows the differences in the structures between the anode single-strip events and the anode multi-strip events. The structures observed in multi-strip events are more abbreviated and exhibit greater curvature compared to those in single-strip events. Specifically, the detected energy on the anode side in multi-strip events tends to decrease near the edges of the structures. This decrease is attributed to charge loss occurring near the gaps between neighboring strips, which is where multi-strip events are typically detected.

Applications of this method that utilize the energy detected by both sides of the detector improved the energy resolutions as demonstrated in Figure \ref{fig:reconst-spectrum-Ba+Na+Co} where the reconstructed energy spectra using all events data, including both single and multiple strip events are shown.
We succeeded in improving energy resolutions. 
The resultant energy resolutions $\Delta E_\mathrm{FWHM}$ were 2.6 keV at 122 keV, 4.3 keV at 356 keV, and 7.2 keV at 511 keV. The improvement in energy resolution, as expected, increases as the energy increases. Under the conditions of the high voltage of 500 V and the temperature of about -20 $^\circ$C, we also confirmed the change of the energy resolution in 20 hours is less than 1 \%. 


\section{Reconstruction of the Depth Of Interaction (DOI)}
\subsection{Modeling theoretical responses of the CdTe-DSD}
\label{subsec:modeling}

\begin{figure}[ht!]
\begin{center}
\includegraphics[width=0.75\hsize]{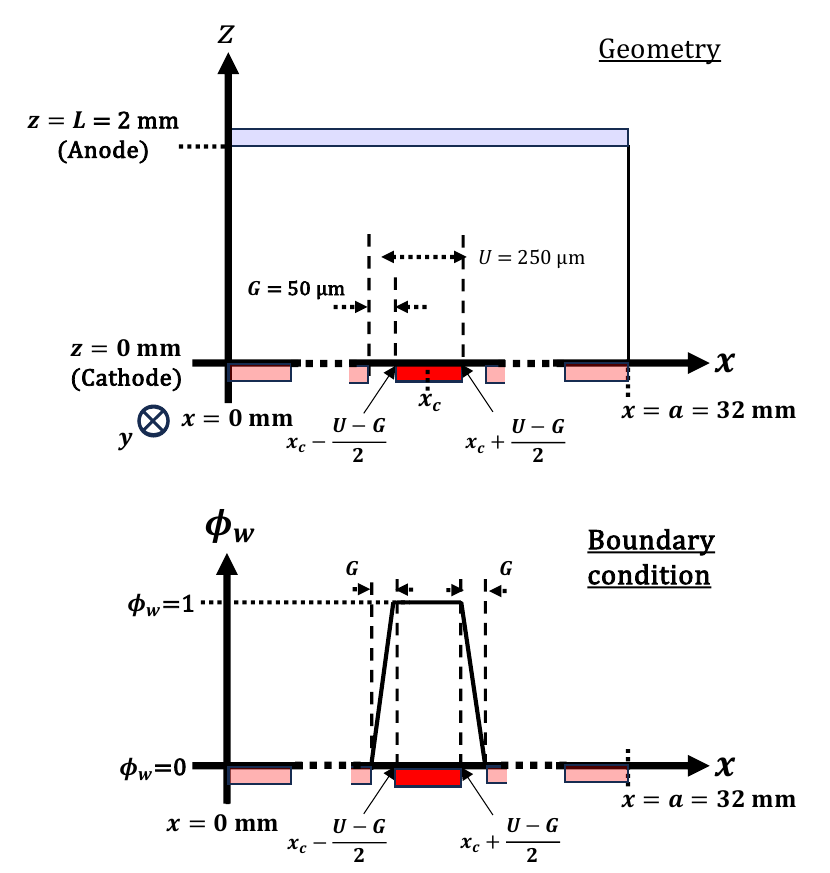}
\caption{The Schematic view of (upper) the geometry of the CdTe-DSD for the simulation and (bottom) the boundary condition of an interest strip on the installed plane.}
\label{fig:cordinate}
\end{center}
\end{figure}

In CdTe detectors, carriers generated by incident photons are not fully collected because of the poor charge transport properties of CdTe. 
In this section, we describe our method of calculating the charge induced on electrodes to correct the incomplete charge collection. Our calculation is based on the Shockley-Ramo theorem and the effect of the mobility-lifetime product ($\mu\tau$) is taken into account \cite{he2001review,hagino2012imaging}.
In the theorem, when charge $q(\Vec{r})$ moves from the interaction position $\Vec{r_i}$ to another position $\Vec{r_f}$, the induced charge $Q$ at an electrode is given by
\begin{equation}
\label{eq:induced_charge_mutau}
Q =-\int_{\Vec{r_i}}^{\Vec{r_f}} q (\Vec{r}) \nabla\phi_w(\Vec{r})\cdot d\Vec{r}, 
\end{equation}
where $\phi_w(\Vec{r})$ is the weighting potential and
 is calculated by solving the Poisson equation $\Delta\phi_w(\Vec{r})=0$ with boundary conditions of $1$ for the readout electrode and $\phi_w(\Vec{r})=0$ for all the other electrodes. The potential is assumed to be a linearly decreasing function from 1 at the edge of the interest electrode to 0 at the nearest edge of the adjacent electrode shown in Figure \ref{fig:cordinate}. 
The weighting potential of the electrode in a double-sided strip detector \cite{NAGASAWA2023168175} is 
\begin{align}
 \label{eq:wp_CdTe-DSD_gap}
 &\mbox{Cathode-side strip:}\notag\\
 &\mbox{ }\phi_w(x,z) = \sum_{m=1}^\infty A_m(x_\mathrm{c})\sin\left(\frac{m\pi}{a}x\right)\sinh\left(\frac{m\pi}{a}z\right),\\
 &\mbox{Anode-side strip: }\notag\\
 &\mbox{ }\phi_w(y,z) = \sum_{m=1}^\infty A_m(y_\mathrm{c}) \sin\left(\frac{m\pi}{a}y\right)\sinh\left(\frac{m\pi}{a}(L-z)\right),\\
 \mbox{}&\notag\\
&A_m(r_\mathrm{sc}) = \frac{8a}{(m\pi)^2G\sinh\left(\frac{m\pi}{a}L\right)}\notag\\
&\times\sin\left(\frac{m\pi}{a}r_\mathrm{sc}\right)\sin\left(\frac{m\pi}{a}\frac{U}{2}\right)\sin\left(\frac{m\pi}{a}\frac{G}{2}\right),\notag
\end{align}
 where $a$ is the detector size, $U$ is the width of the strip pitch, $G$ is the width of a gap, and $L$ is the thickness of the detector. The center of an interest strip electrode is $x=x_c, z=0$ for the cathode side and $y=y_c, z=L$ for the anode side shown in Figure \ref{fig:cordinate}. In addition, we assumed the center position of the electrode $x_c$, $y_c$ for each side is the center of the detector $\frac{a}{2}$, where the detector response could be regarded as the average one of all the strips. 

For the sake of simplicity, we assumed that the charge cloud is point-like at the initial stage, that the electric field $E$ is parallel, and that there is no diffusion. Under these conditions, the carrier velocity $v=\mu E$ was constant, and the charge $q_\mathrm{e,h}$ of each carrier produced by an incident photon was given by
\begin{equation}
\label{eq:carrier_time_dep}
q_\mathrm{e,h}(t)=(\pm e)n\exp\left(-\frac{|z-z_i|},{(\mu\tau)_\mathrm{e,h}E}\right),
\end{equation}
where $n$ is the number of carriers, which is proportional to the incident energy in the keV band, and $z_i$ is the DOI.
If we set the position of the cathode side at z=0, the electric field in CdTe is 
\begin{align}
\label{eq:electric field}
E(z)= -\frac{en_\mathrm{sp}}{\epsilon_\mathrm{CdTe}}\left(z-\frac{L}{2}\right)+\frac{H.V.}{L},
\end{align}
where $\epsilon_\mathrm{CdTe}$ is the permittivity of CdTe, $n_\mathrm{sp}$ is the density of the space charge of ionized acceptors, and $H.V.$ is the applied voltage.
The space charge could be negative \cite{Toyama_2006,Cola2013}, and all the volume of the CdTe crystal is depleted when a bias voltage is applied.
Finally, we calculated the induced charges $Q$ on the anode and cathode by using the equations (\ref{eq:induced_charge_mutau}-\ref{eq:electric field}). We also calculated the charge collection efficiency on both sides, dividing the induced charges $Q$ by the product of the number of electron-hole pairs and the elementary charge.

\subsection{Reproducing the DOI dependence of the experimental data using the theoretical detector response}
\begin{figure*}[ht!]
\begin{center}
\includegraphics[width=0.80\hsize]{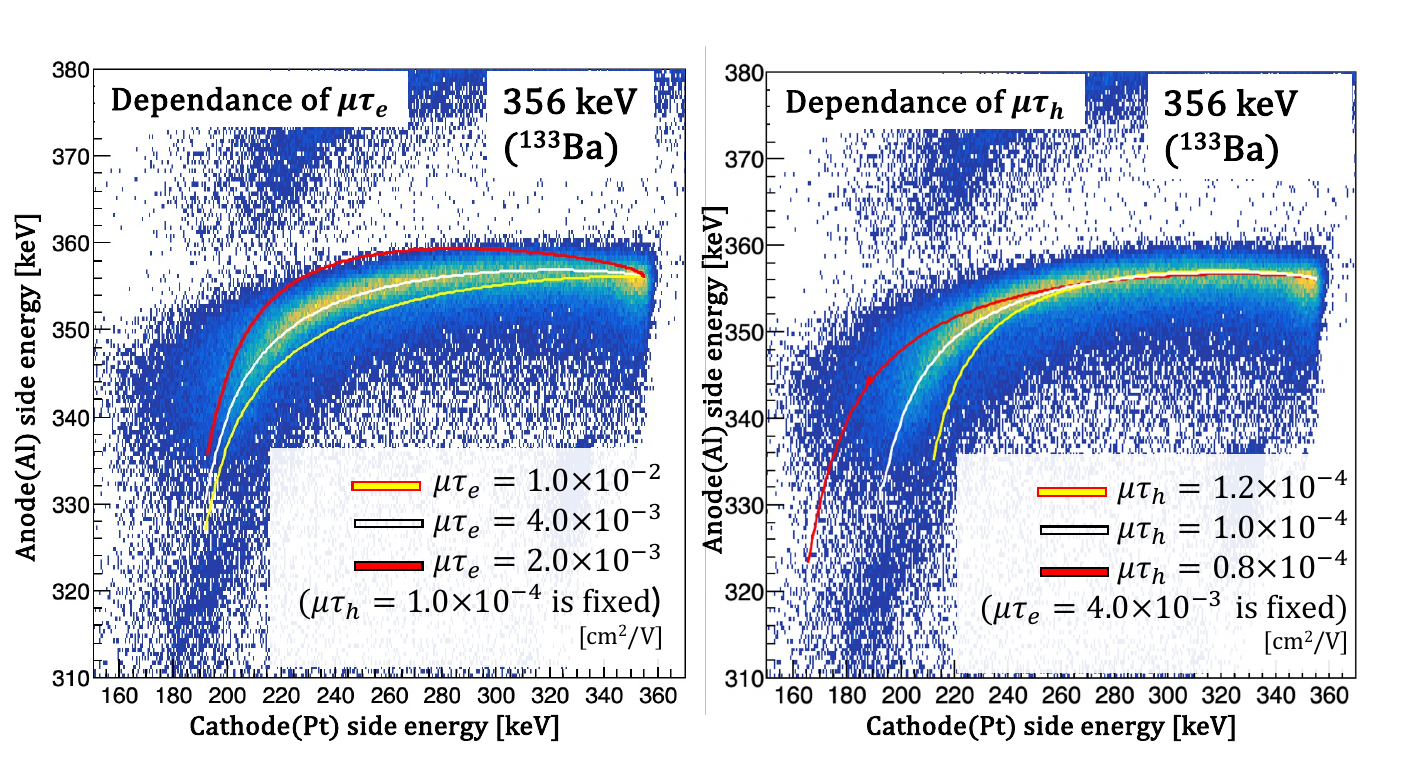}
\caption{Comparison between theoretical curves with varying $\mu\tau$ values for $n_\mathrm{sp} = 0$ and the extended structure whose incident photon energy is 356 keV in single-strip events from experiments. The points for normalization were set at the highest counts in the structure, corresponding to the peak of the energy spectrum.}
\label{fig:theoretical_calc_mutau_dependance}
\end{center}
\end{figure*}
We attempted to reproduce the extended structure in the anode-cathode detected energy
correlation map using a model incorporating mobility and lifetime. 
For arbitrary three-dimensional interaction positions, we calculated the charge collection efficiency of each strip on the anode and cathode sides. The efficiency multiplied by an incident photon energy is the theoretical detected energy when the photon interacts with the detector at a given position. Thus, we reproduced the relationship between detected energies on the cathode and the anode side. 

To make comparisons between the calculation result of the theoretical modeling and the experiment data, we regarded the sum of energies of the two strips with the first and second highest detected energies among the strips with significant signals as the cathode-side energy to reduce the effect caused by the difference in the energy threshold, and hereafter refer to it as the ``cathode1st+2nd energy''. The cathode1st+2nd energy is considered to depend on DOI significantly and be nearly independent of the 2D position from the simulation results of our model. For the anode-side energy, we selected single-strip events only (i.e., a type of event where only one strip shows a signal significantly over the threshold for an incoming photon), of which we considered that the interaction position of the photon with CdTe is near the center of a strip and the diameter of the diffused charge cloud is smaller than the width of the strip. In addition, we also considered the detector response does not depend on the interaction position along the strip (the direction of the strip width) if the interaction position is inside the strip. Therefore, we consider the anode single-strip events are observed when their interaction positions are near the strip center, which is consistent with the assumptions of the parallel electric field and point-like initial charge cloud. 
Thus, the cathode1st+2nd energy in the anode single-strip events is optimal for comparing our model and the experimental data.

To determine the parameter values of $\mu\tau_\mathrm{e,h}$, we compared the measured extended structure whose incident photon energy is 356 keV and the result of the theoretical modeling.
Figure \ref{fig:theoretical_calc_mutau_dependance} shows the result of a comparison between theoretical curves with varying $\mu\tau$ values for $n_\mathrm{sp} = 0$ and the structure of 356 keV in single-strip events from experiments. 
We set the step size of $\Delta\mu\tau_\mathrm{e,h}$ value to $1\times 10^{-3}, 5\times 10^{-6}$ cm$^2$/V, respectively. We calculated a difference between a center of peak in a slice of the structure at an arbitrary $x$ (cathode1st+2nd) and the $y$ (anode energy) of the theoretical curves at an arbitrary $x$ (cathode1st+2nd) shown in Figure \ref{fig:theoretical_calc_mutau_dependance} and summed the squared differences varying x values (cathode1st+2nd). 
While changing the $\mu\tau_{e,h}$ values with intervals of the $\Delta\mu\tau_{e,h}$ independently, we determined the $\mu\tau$ values as the best-matched parameter for a certain space charge with accuracy of $\Delta\mu\tau_{e,h}$ when the sum is the smallest.
The best-matched parametersl were $\mu\tau_e = 4 \times 10^{-3}$, $\mu\tau_h = 1.0 \times 10^{-4} $ cm$^2$/V for $n_\mathrm{sp}=0.0$. We also performed comparisons for $n_\mathrm{sp}=-6.0$ and $-11\times10^{10}$ cm$^{-3}$ and list the best-matched values in Table \ref{tb:optimal_mutau}. 
\begin{table}[ht!]
\caption{Space charge density and optimal $\mu\tau$ values}
\begin{center}
\begin{tabular}{ c|c c }
 $n_\mathrm{sp}$ [/cm$^3$] & $\mu\tau_e $ [cm$^2$/V]& $\mu\tau_h$ [cm$^2$/V]\\ \hline\hline
 0.0 & $4 \times 10^{-3}$ & $ 1.0 \times 10^{-4} $ \\ 
 $-6 \times 10^{10}$ & $4 \times 10^{-3}$ & $ 1.15 \times 10^{-4} $\\
 $-11 \times 10^{10}$ & $5 \times 10^{-3}$ & $ 1.15 \times 10^{-4} $
\end{tabular}
\label{tb:optimal_mutau}
\end{center}
\end{table}

\subsection{Reconstruction of Depth Of Interaction(DOI)}

\begin{figure*}[h]
\begin{center}
\includegraphics[width=0.475\hsize]{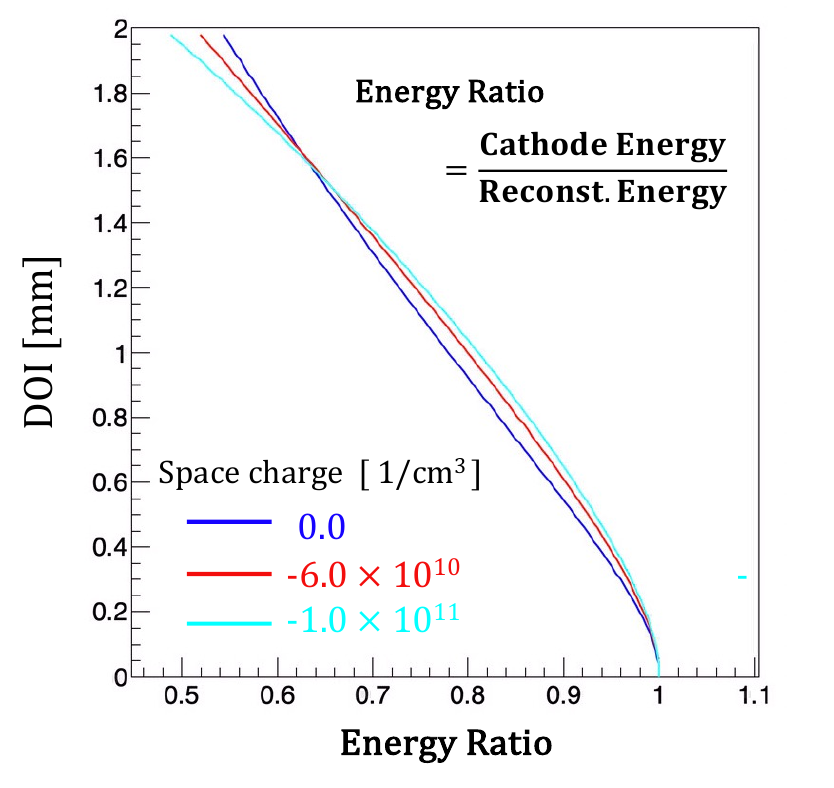}
\caption{ Theoretically-calculated functions to determine the DOI from a DOI index\ (the ratio of the cathode1st+2nd energy (see text) to the reconstructed energy) for each space charge density shown in Table \ref{tb:optimal_mutau}.}
\label{fig:functions}
\end{center}
\end{figure*}
\begin{figure*}[ht!]
\begin{center}
\includegraphics[width=0.9\hsize]{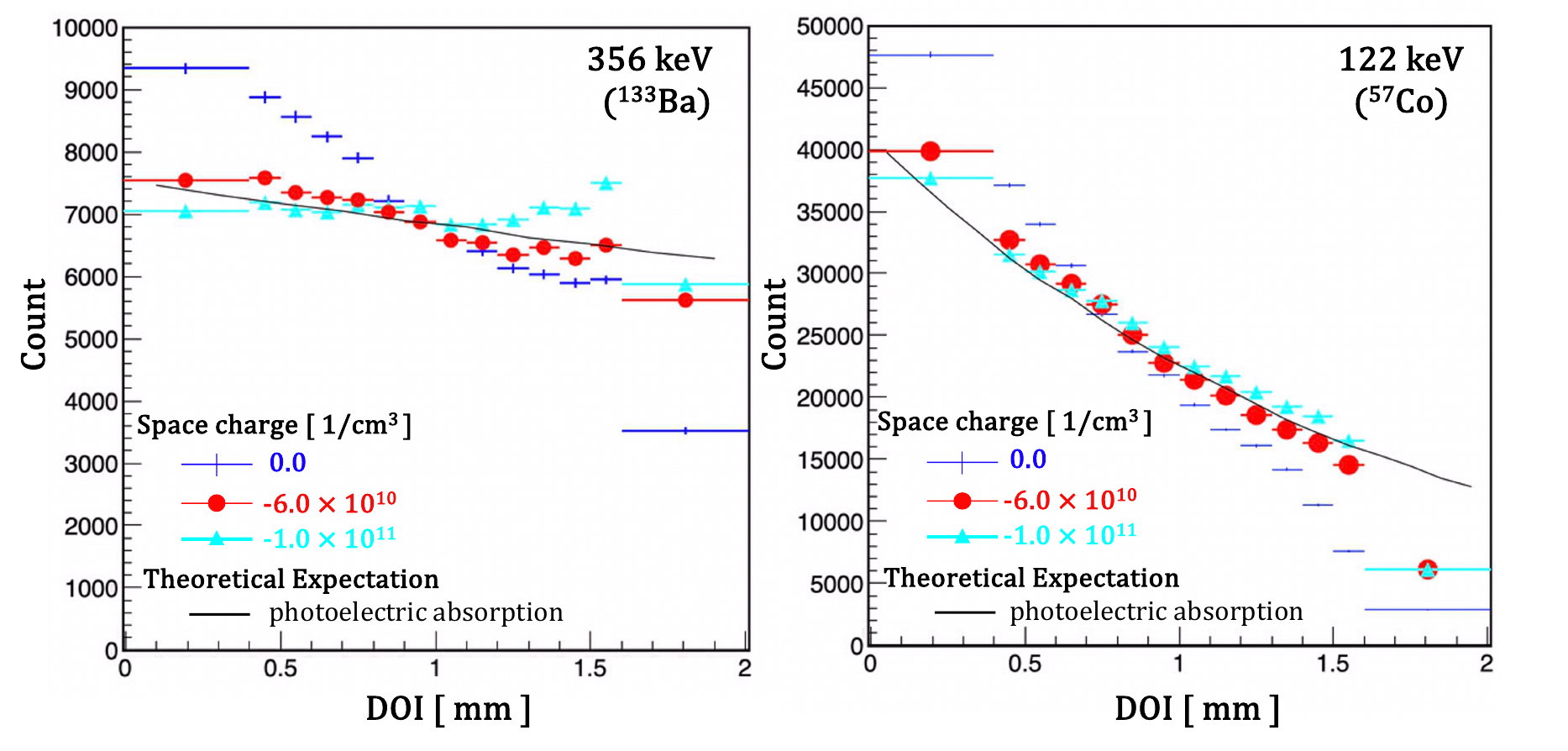}
\caption{Reconstructed DOIs for each $n_\mathrm{sp}$ and the theoretical expectations of (right) 122 keV from $^{57}$Co and (left) 356 keV from $^{133}$Ba. The points are the result of the data, and the black line is the simulated result with the Geant4 simulation tool kit. }
\label{fig:zdepth}
\end{center}
\end{figure*}

We created functions from a theoretical relationship between the DOI and a certain observable quantity. Given experimental data corresponded to the observable quantity, the function returns the DOI.
As an observable quantity, we selected the induced charge efficiency of cathode1st+2nd using our response model, corresponding to the ratio of the cathode1st+2nd energy to the reconstructed energy in the experimental data because the reconstructed energy does not vary with the DOI and can be treated as a normalization factor.
The theoretical efficiency of cathode1st+2nd varies significantly with the DOI, and the experimental ratio can also be very sensitive to the DOI.
In addition, the reconstructed energy and the cathode1st+2nd energy are experimentally less susceptible to the electrical noise of the strip or the given event threshold because they are sufficiently higher than the values of the noise or the threshold. 
Figure \ref{fig:functions} shows the functions to estimate the DOI from the ratio of the cathode1st+2nd energy to the reconstructed energy. 
 
We applied formulated functions to the experimental data. Figure \ref{fig:zdepth} shows the result for events for 122 keV from $^{57}$Co and 356 keV from $^{133}$Ba for the three selected space charge densities (Table~\ref{tb:optimal_mutau}) and the theoretical expectations. We extracted the events of the experimental data around each peak in the range of 3 $\Delta E_\mathrm{FWHM}$ of the energy resolution shown in Figure \ref{fig:reconst-spectrum-Ba+Na+Co}. We performed a simulation for the theoretical expectations, using the Geant4 simulation tool kit \cite{geant4} to obtain the reaction positions for events where only photoelectric absorption occurred. 

The reconstructed DOIs were found to be consistent with the theoretical expectations at both 122 keV and 356 keV for $n_\mathrm{sp}=-6.0\times10^{10}$ cm$^{-3}$. 
The values of $\mu\tau_h$ and $\mu\tau_e$ are consistent with results reported in the literatures \cite{Estrada_2014}.
The derived space-charge density is almost consistent with the homogeneous distribution of all ionized acceptors except for the deep-lying acceptors as reported in the literature \cite{6202392}.

Our finding implies that the DOI can be determined with an accuracy of about 100 $\mu$m except for those at the edges of the detector. 
The accuracy is determined from the energy resolutions of reconstructed energy and cathode1st+2nd energy (or the breadth of the extended structure for arbitrary energy of the incident photon). 
However, there is a significant difference between the results of 122 keV and 356 keV for $n_\mathrm{sp}=-6.0\times10^{-10}$ cm$^{-3}$ and the theoretical expectations at $z = 2.0$ mm. 
We attributed this difference most likely to further spreading around the left edge of the extended structure shown in Figure \ref{fig:Corerala+both_side_spectrum_133Ba-2} or \ref{fig:theoretical_calc_mutau_dependance}.
The spreading implies poor energy resolution, resulting in reduced sensitivity to DOI.

\section{Imaging Performance and Response Uniformity of the CdTe-DSD }
\subsection{Imaging Performance}
We evaluated the imaging performance of the CdTe-DSD with a test target of a 0.1-mm-thick lead (Pb) X-ray test chart, which accommodates 8 different interval pitches between the apertures and masks. The widths of the apertures and masks are the same and are equal to a half of the interval pitches. The experiment was set up as follows. The distances of the detection surface to the test chart and to a radioisotope $^{133}$Ba source, whose diameter is $\sim$ 
 10 mm, were 30 and 330 mm, respectively. The sources were small and sufficiently far apart to the extent that the width of the aperture and that in the image at the detection plane was approximately the same. We used X-ray lines of 30.6 and 31.0 keV from the $^{133}$Ba source, for which almost all photons are stopped by the 0.1-mm-thick lead (Pb).

We reconstructed the image, assuming that the incident position of the photon at each strip follows a Gaussian probability distribution \cite{furukawa2020imaging} with
the mean position being at the center of the strip. The sigma of the Gaussian was calculated with the percentage of the charge-shared events; the percentage of single-strip events at $\sim$31 keV were about 70\% and 80\% for the anode and cathode sides, respectively, and accordingly, the two sigmas were 250 $\mu$m$\times$ 70\% = 175 $\mu$m and 250
$\mu$m$\times$ 80\% = 200 $\mu$m. 

\begin{figure}[ht!]

\begin{center}
\includegraphics[width=\hsize]{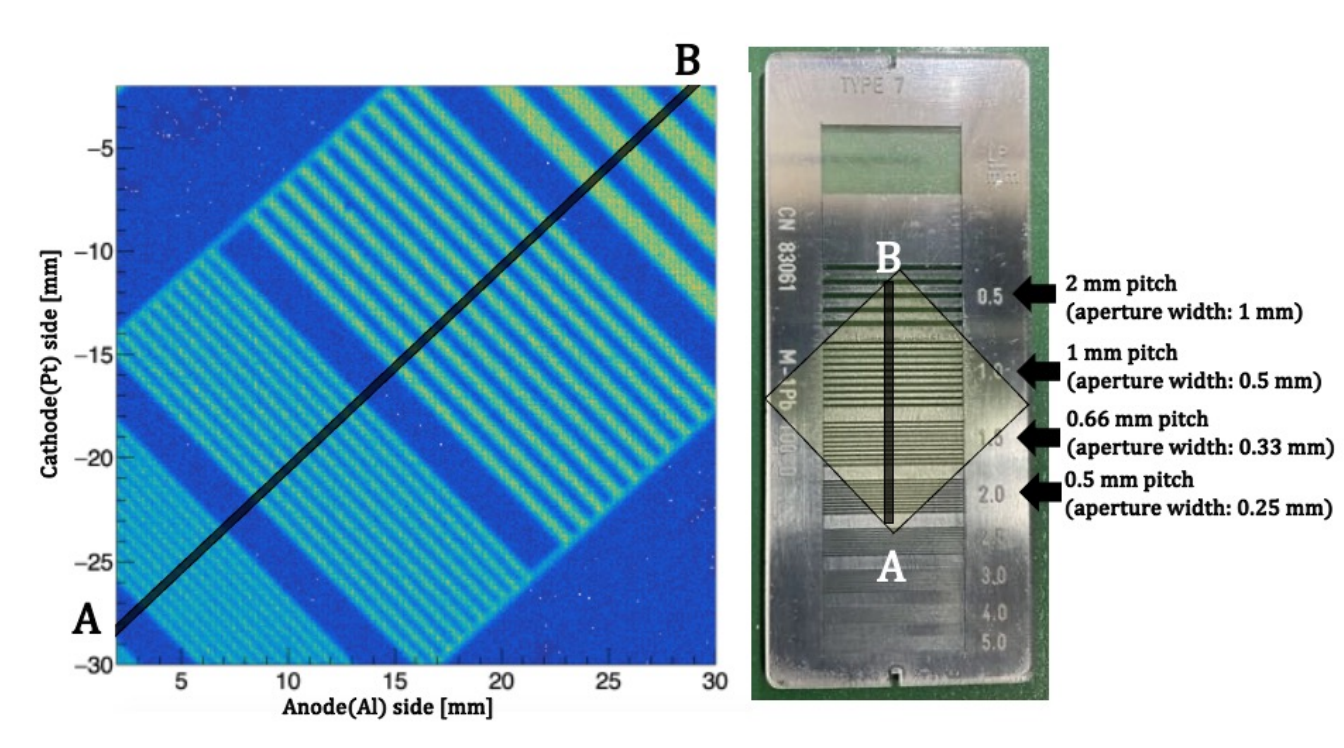}
\includegraphics[width=\hsize]{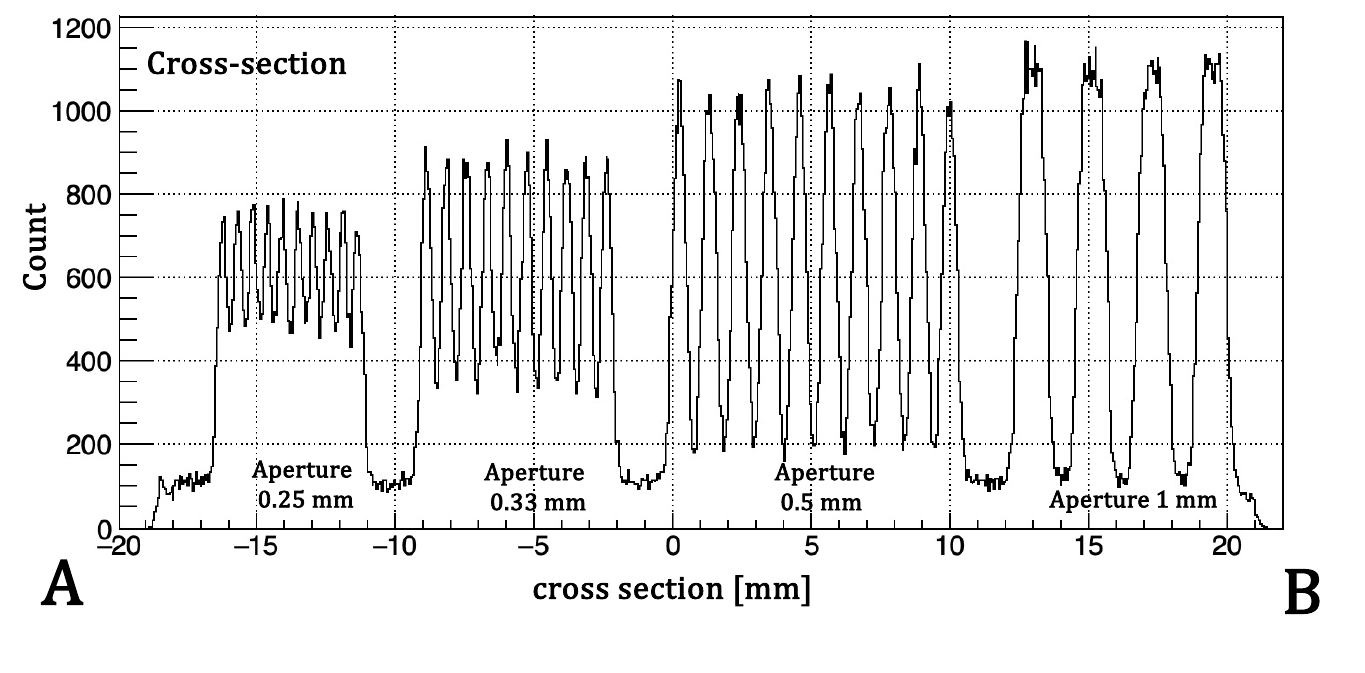}
\caption{ (upper right) A picture of a 0.1-mm-thick lead (Pb) X-ray test chart, (upper left) an acquired image with 0.5, 0.66, 1.0, and 2.0-mm pitch of the chart using the 31-keV-X-ray line from $^{133}$Ba. (lower panel) Cross-section profile from points A to B on the image (upper-right). The yellow square on the picture shows the imaging area. }
\label{fig:image_positional_resolution}
\end{center}
\end{figure}

Figure \ref{fig:image_positional_resolution} shows a photograph of the test chart and the acquired image in combination with 0.5, 0.66, 1.0, and 2.0 mm pitches of the slits and its cross-section profile. We confirmed that a structure of 250 $\mu$m can be resolved, as expected from the design value of 250 $\mu$m for the spatial resolution of the CdTe-DSD.

\subsection{Uniformity of the CdTe-DSD}
We evaluated the potential positional variances of the detection efficiency and energy resolution of the CdTe-DSD. To make quantitative measurements in various applications, uniformity across the entire detector is a critical element, as well as the high spatial resolution.
Since the edge areas are affected by electrical noise anyway, the events in the central 30 $\times$ 30 mm$^2$ area were used for evaluation.

The experiment was set up as follows. The distance of the detection surface to a radioisotope was $\sim$ 300 mm. An error for the number of incident photons due to the difference in length $d$ from the source to the detection surface was estimated to be $(a/d)^2 = (30/300)^2 \sim 1 \% $, where $a = 30$ mm is the size of the detection area. We used an X-ray line of 122.1 ($\pm 3 \Delta E_\mathrm{FWHM}$) keV from a $^{57}$Co source whose average penetration was ~ 2 mm, and therefore the photons from the source hit the entire CdTe diode sensitive volume. 

\begin{figure}[ht!]
\begin{center}
\includegraphics[width=0.95\hsize]{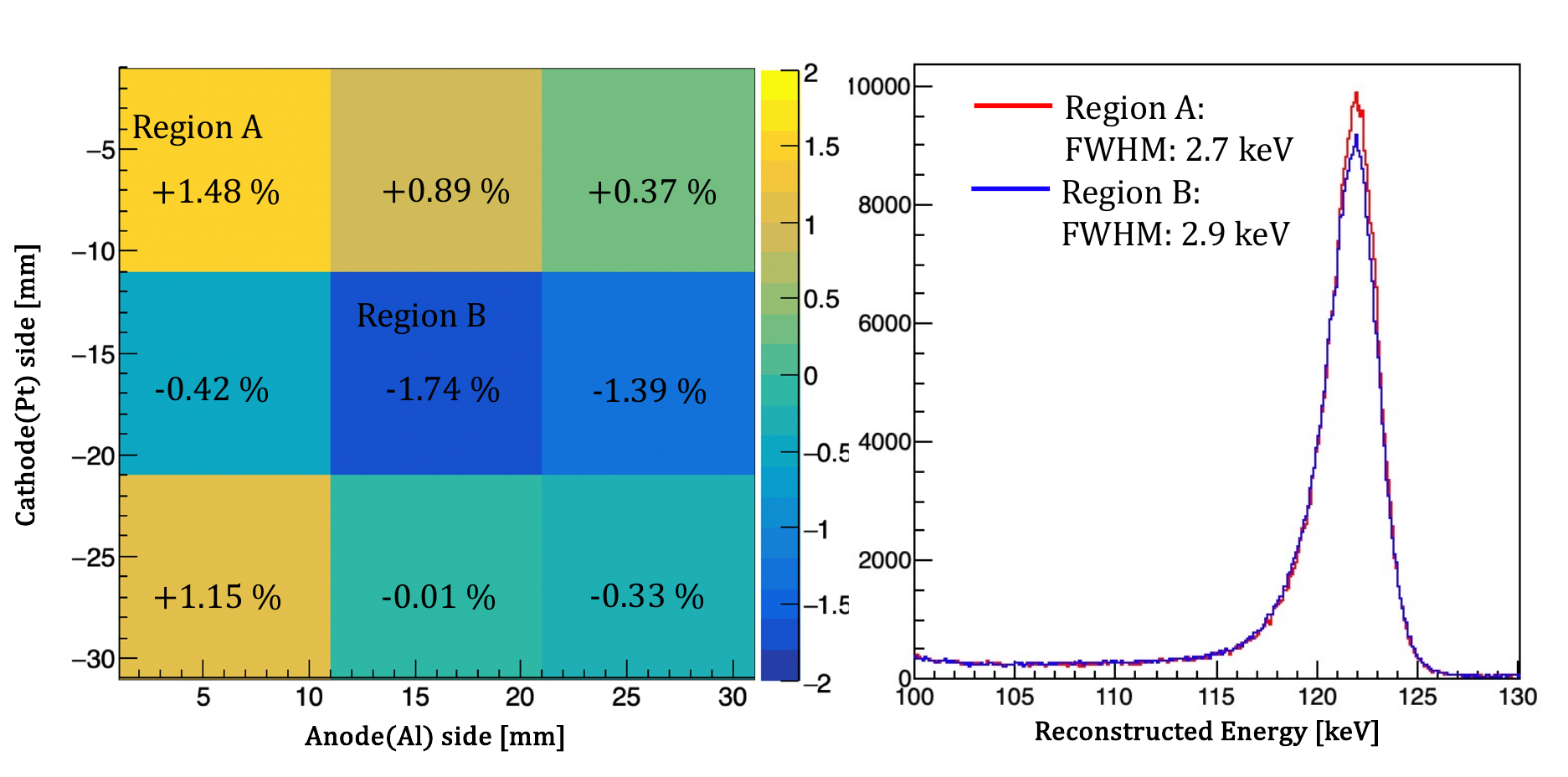}
\caption{ (left) Differences between the integrated values in the range of 122.1 $\pm 3 \times 2.8$ keV ($=3\Delta E_\mathrm{FWHM}$ shown in Figure \ref{fig:reconst-spectrum-Ba+Na+Co}) in nine divided detection regions (normalized by the average value of the whole integrated values). (right) Comparison of the spectra with the maximum\ (Region A, red line) and minimum counts\ (Region B, blue line). The respective spectral resolutions are 2.7 and 2.9 keV (FWHM).}
\label{fig:uniformity}
\end{center}
\end{figure}

Figure \ref{fig:uniformity} shows the percentage difference between the integrated value of the peak at 122.1 keV in each of the nine divided detection regions (normalized by the average counts of the whole integrated values) and the comparison of the spectra with the maximum and minimum counts among the sample. The standard deviation of the counts in each region was about 1\%, and the statistical variance of the average value is $\sim$0.1\%, which is negligible compared with the standard deviation of the counts. The standard deviation is comparable to the estimated error. In other words, the positional-dependent variation across the detector is less than 1\%. 
The energy resolutions of the spectra from the two regions with the maximum and minimum counts (designated as regions A and B, respectively, in Figure \ref{fig:uniformity} and of the overall spectrum were 2.7, 2.9, and 2.8 keV, respectively, which could be caused by differences in the wiring length from CdTe-DSD to ASICs. 
Considering that there was no significant positional variation in the energy resolutions, we concluded that our energy reconstruction was appropriate regardless of the regions and guaranteed a good spectroscopic response uniformity.

\section{Conclusion}
We have developed a 2-mm-thick CdTe double-sided strip detector (CdTe-DSD). The detector has high spatial resolution with a uniform large area and high energy resolution with high detection efficiency in tens to hundreds of keV. The detector was successfully operated at a temperature of -20 $^\circ$C and with an applied bias voltage of 500 V, and its energy resolution is stable in 20 hours.
The increased thickness exacerbated the poor carrier transport properties, which is an intrinsic weak point of the CdTe. The detected energy depends on the depth of interaction (DOI).

We reconstructed the low-energy tails that exacerbate the loss in the energy resolutions from information obtained from both the cathode and anode sides of the detector and determined an energy resolution to be 4.3 keV at 356 keV. 
We also developed a theoretical model of the detector response.
Using it, we derived the optimal mobility-lifetime product\ ($\mu\tau$) of carriers $\mu\tau_{e,h}=(4 \pm 1 )\times 10^{-3}, (1.15\pm 0.05) \times 10^{-4}$ [cm$^2$/V] with the space charge density of $n_{sp}=-6.0\times 10^{10}$ [1/cm$^{-3}$] in this CdTe detector and reconstructed the DOIs for 122 keV and 356 keV incoming photons with an accuracy of 100 $\mu m$ using the ratio of the significant detected energy on the cathode side (cathode1st+2nd) to the reconstructed energy. The results were consistent with their expected penetration depths. 
For the DSD detectors, our reconstruction method of DOI using $E_\mathrm{cathode1st+2nd}/E_\mathrm{reconst}$ is more sensitive to DOI than the well-known method of $E_\mathrm{cathode}/E_\mathrm{anode}$.
Both induced charge efficiencies on the anode and the cathode side in the $E_\mathrm{cathode}/E_\mathrm{anode}$ ratio decrease when the DOI is around the anode surface because the hole contribution and the effect of hole trapping become larger simultaneously, resulting in low sensitivity to the DOI near the anode surface. On the other hand, the $E_\mathrm{cathode1st+2nd}$ significantly depends on the DOI, and the $E_\mathrm{reconst}$ is the normalization factor.

We also evaluated the spatial resolution demonstrating to resolve patterns of 250-$\mu$m-width slits in an X-ray test chart and the positional-dependent variation across the detector to be smaller than 1\%. 

The 2 mm thick CdTe DSD we have developed with the reconstruction method of the incident photon energy and the DOI, resulting in acquiring high-resolution information of energy and three-dimensional interaction positions with uniform large imaging area, is optimal for quantitative observation of hard X-ray and soft gamma-ray with spectroscopic imaging. To realize more detailed imaging, the compact readout due to the DSD configuration is an advantage, and our new methods
could be applicable to thick double-sided strip detectors with fine-pitch strips of CdTe or CdZnTe.

\section*{Acknowledgment}
This work was supported by JSPS, Japan KAKENHI Grant Numbers 21H00163, 21K18049, 
22KJ0873 and 22J12583. 
S.N. is also supported by FoPM (WINGS Program) and JSR Fellowship, the University of Tokyo, Japan.
The authors wish to thank Olivier Limousin for fruitful discussions.

\bibliography{nima}
\end{document}